\documentclass[superscriptaddress,nofootinbib, amsmath,amssymb,preprintnumbers,prdfloatfix,twocolumn,PRD]{revtex4-2}
\usepackage{CJK}
\usepackage{graphicx}
\usepackage{dcolumn}
\usepackage{upgreek}
\usepackage{slashed}
\usepackage{braket}
\usepackage{amsmath}
\usepackage{bm}
\usepackage[colorlinks=true,pdfstartview=FitV,breaklinks=true]{hyperref}
\usepackage[dvipsnames,table]{xcolor}
\hypersetup{urlcolor=BlueViolet,
	    citecolor=Plum,
	    linkcolor=PineGreen}
\usepackage{lipsum}	    
\usepackage{titlesec}
\usepackage{etoolbox}
\usepackage[normalem]{ulem}
\usepackage{subcaption}
 \usepackage{lettrine,Typocaps}

\usepackage{relsize}

\usepackage{slashed}

\newcommand{\htwo}{\mathrm{H_2}}
\newcommand{\hminus}{\mathrm{H^-}}

\newcommand{\cchi}{\mathcal{X}}

\newcommand{\equaref}[1]{Eq.~(\ref{#1})}

\newcommand{\figref}[1]{Fig.~\ref{#1}}

\newcommand{\secref}[1]{Section~\ref{#1}}

\newcommand{\tabref}[1]{Table~\ref{#1}}
\newcommand{\refref}[1]{Ref.~\cite{#1}}
\newcommand{\refsref}[2]{Refs.~\cite{#1}~and~\cite{#2}}

\definecolor{offblue}{RGB}{23,80,153}

\begin{document}

\preprint{IPMU25-0034}

\title{Neutrino masses, matter-antimatter asymmetry, dark matter, and supermassive black hole formation explained with Majorons}

\author{Yifan Lu}
\email{yifanlu@ucla.edu}
\affiliation{Department of Physics and Astronomy, University of California Los Angeles,\\ Los Angeles, California, 90095-1547, USA}

\author{Zachary S. C. Picker} 
\email{zpicker@physics.ucla.edu}
\affiliation{Department of Physics and Astronomy, University of California Los Angeles,\\ Los Angeles, California, 90095-1547, USA}

\author{Alexander Kusenko}
\email{kusenko@ucla.edu}
\affiliation{Department of Physics and Astronomy, University of California Los Angeles,\\ Los Angeles, California, 90095-1547, USA}
\affiliation{Kavli Institute for the Physics and Mathematics of the Universe (WPI), UTIAS \\The University of Tokyo, Kashiwa, Chiba 277-8583, Japan} 

\author{Tsutomu T. Yanagida}
\email{tsutomu.tyanagida@sjtu.edu.cn}
\affiliation{Kavli Institute for the Physics and Mathematics of the Universe (WPI), UTIAS \\The University of Tokyo, Kashiwa, Chiba 277-8583, Japan}
\affiliation{Tsung-Dao Lee Institute \& School of Physics and Astronomy, \\Shanghai Jiao Tong University, Pudong New Area, Shanghai 201210, China}

\begin{abstract}
\noindent The spontaneous breaking of a global lepton number symmetry can result in a (pseudo) Nambu-Goldstone boson known as the Majoron. We study a singlet Majoron model that couples to two Higgs doublets in which the lepton number current develops an electromagnetic anomaly, allowing the decay of Majorons into photons. We focus on Majorons at the eV scale with an enhanced anomaly and show that it serves as a dark matter candidate whose decay signals can be probed by space telescope observations. Furthermore, if the decay produces Lyman-Werner photons, heavy black hole seeds can be generated via the direct collapse mechanism and evolve into the active galactic nuclei we observe at high redshifts. Our framework thus simultaneously addresses the origin of neutrino masses, the baryon asymmetry of the Universe, the nature of dark matter, and the formation of high redshift supermassive black holes.
\end{abstract}

\maketitle

\section{Introduction}

\noindent A number of important questions are not explained by the Standard Model of particle physics (SM). These include the origin of neutrino mass, the production of the observed baryon asymmetry of the Universe, and the nature of dark matter. One well-motivated solution to the origin of the neutrino masses involves extending the Standard Model to include a spontaneously broken global lepton number symmetry $U(1)_L$. If the lepton number symmetry is broken by the vacuum expectation value (VEV) of a singlet scalar at a high scale, right-handed neutrinos acquire large Majorana masses, which can `seesaw' the left-handed neutrinos, forcing them to acquire naturally light masses \cite{Minkowski:1977sc, Yanagida:1979as, Yanagida:1979gs, Glashow:1979nm, Gell-Mann:1979vob}. In the theory of \textit{leptogenesis} \cite{Fukugita:1986hr, Buchmuller:2005eh}, the CP-violating decay of these right handed-neutrinos in the early Universe is then able to generate a lepton asymmetry. This is then converted to a baryon asymmetry through sphaleron processes \cite{Kuzmin:1985mm}. The angular mode of the singlet scalar gives rise to a pseudo Nambu-Goldstone boson known as the Majoron \cite{Chikashige:1980qk, Chikashige:1980ui, Gelmini:1980re}.

In this unified framework, it is even more appealing to consider the possibility that the Majoron is a dark matter candidate \cite{Rothstein:1992rh,  Langacker:1986rj, Lattanzi:2007ux, Gu:2010ys, Frigerio:2011in, Wang:2016vfj, Abe:2020dut, Manna:2022gwn, Berezinsky:1993fm, Bazzocchi:2008fh, Queiroz:2014yna, Dudas:2014bca, Garcia-Cely:2017oco, Greljo:2025suh}. When the Majoron is heavy ($>\mathrm{keV}$) and interacts weakly or feebly, it can be produced via freeze-out or freeze-in \cite{Frigerio:2011in}. Such a scenario could be probed by the emission of neutrino lines, X-rays and gamma-rays \cite{Berezinsky:1993fm, Bazzocchi:2008fh, Queiroz:2014yna, Dudas:2014bca, Garcia-Cely:2017oco, Lattanzi:2013uza, Akita:2023qiz}.

There is, however, another possibility---the Majoron could be a light ($\lesssim~\mathrm{eV}$) particle, produced via the misalignment mechanism after the spontaneous breaking of the lepton number symmetry. In this case, the detectability of the Majoron relies on the existence of an electromagnetic anomaly that allows its decay into low energy photons. Recently, \refref{Liang:2024vnd} proposed a light Majoron model where the $U(1)_L$ charge assignment admits a Majoron-photon coupling, so that current axion haloscopes could be used to probe $\mathrm{\mu eV}$ Majoron masses. 

However, the mass window that lies in the eV mass range may be even more interesting. As we will show in this work, eV-scale Majoron dark matter may be naturally favored due to considerations from both the generation of the neutrino masses along with the misalignment abundance. Even more intriguingly, the decay of eV-scale Majorons may be detectable with existing or future space telescopes, such as the James Webb Space Telescope (JWST) and the Hubble Space Telescope (HST). Photons from Majoron decay in the infrared, optical and ultraviolet regimes allow us to constrain these Majoron models using spectral line searches \cite{Todarello:2023hdk,Todarello:2024qci,Pinetti:2025owq} or from its impact on the cosmic optical background anisotropy \cite{Carenza:2023qxh}. Remarkably, the current JWST and HST data can already constrain the Majoron photon coupling down to $\sim 10^{-12}~\mathrm{GeV}^{-1}$, with the possibility of improvements in the infrared range as JWST observations improve.

In this work, we will consider Majoron models with large electromagnetic anomalies. The resulting enhancement in the Majoron decay rate offers a promising target for detection by current astrophysical instruments. Furthermore, we will show that for Majorons that decay into Lyman-Werner (LW) photons, the radiation background provides the necessary conditions within pristine protogalaxies for the rapid assembly of supermassive black holes (SMBHs) via the so called `direct collapse' process \cite{Lu:2024zwa,Lu:2023xoi,Omukai:2000ic, Oh:2001ex, Begelman:2006db, Inayoshi:2019fun, Bromm:2002hb, Choi:2013kia, shang2010supermassive, Dijkstra:2008jk, becerra2018assembly}. This offers a possible explanation to the origin of the mysterious population of very high redshift SMBHs being discovered by JWST \cite{Bogdan:2023ilu, ubler2023ga, larson2023ceers, harikane2023jwst, carnall2023massive, onoue2023candidate, kocevski2023hidden, Maiolino:2023zdu, fan2023quasars}. Our work is therefore able to simultaneously explain four outstanding, major questions across astrophysics and particle physics---the nature of dark matter, the origin of the neutrino masses, the origin of the matter/antimatter asymmetry, and the unexpected early formation of supermassive black holes.

\begin{table*}[ht!]
\centering
\begin{tabular}{ | c || c | c | c | c | c | c | c | c | c| } 
 \hline
 Fields & $q_L$ & $u_R$ & $d_R$ & $l_L$ & $e_R$ & $N_R$ & $H_1$ & $H_2$ & $\phi$ \\ 
 \hline
 $U(1)_{EM}$ & $(2/3, -1/3)$ & $2/3$ & $-1/3$ & $(0, -1)$ & $-1$ & $0$ & $(1, 0)$ & $(1, 0)$ & $0$ \\ 
 \hline
 $U(1)_L$ & $\mathcal{X}_q$ & $\mathcal{X}_q + \mathcal{X}_1$ & $\mathcal{X}_q - \mathcal{X}_1$ & $1/2 - \mathcal{X}_1$ & $1/2 - \mathcal{X}_1 - \mathcal{X}_2$ & $1/2$ & $\mathcal{X}_1$ & $\mathcal{X}_2$ & $1$\\ 
 \hline
\end{tabular}
\caption{Field content and charge assignment of the two Higgs doublet Majoron model.}
\label{charges}
\end{table*}

The paper is organized as follows: In \secref{Majoronmodel}, we introduce the Majoron model with two Higgs doublets and an enhanced electromagnetic anomaly. In \secref{Majorondm}, we discuss the misalignment production of Majorons and the existing astrophysics constraints. We review the direct collapse mechanism and identify the viable parameter space in \secref{SMBH}, and we draw our conclusions in \secref{conclusion}.

\section{The Two Higgs Doublet Majoron Model}
\label{Majoronmodel}

\noindent We begin with the Majoron model proposed in \refref{Liang:2024vnd}, where we introduce two Higgs doublets $H_1$, $H_2$, and a single scalar $\phi$ carrying the $U(1)_L$ charge that couples to right handed neutrinos $N_R$. The spontaneous symmetry breaking of $U(1)_L$ through the VEV $\braket{\phi}$ at a high scale induces a heavy Majorana mass for $N_R$ and a Nambu-Goldstone boson $J$, the Majoron \cite{Chikashige:1980qk, Chikashige:1980ui, Gelmini:1980re}. Specifically, the Yukawa sector of the model is given by
\begin{equation}
\begin{aligned}
\mathcal{L} & \supset \bar{q}_L Y_u u_R \tilde{H}_1+\bar{q}_L Y_d d_R H_1+\bar{\ell}_L Y_e e_R H_2 \\
& +\bar{\ell}_L Y_D N_R \tilde{H}_1+ \frac{1}{2}\bar{N}_R^c Y_N N_R \phi^* +\text { h.c. } .
\end{aligned}
\label{model}
\end{equation}
It is worth noting that the quark sector only couples to $H_1$, and the charged leptons and the neutrinos separately couple to $H_2$ and $H_1$. Thus, this model does not induce flavor changing neutral currents (FCNCs). The Yukawa couplings do not entirely fix the lepton numbers. Once the U$(1)_L$ charge of $\phi$ is chosen to be $\cchi_{\phi} = 1$, the remaining degrees of freedom in the assignment of $U(1)_L$ charge can be parametrized in the charges of left-handed quarks ($\cchi_{q}$), and the charges of the two Higgs ($\cchi_1$ and $\cchi_2$). The electromagnetic and the $U(1)_L$ charge assignment of the model are outlined in \tabref{charges}. This \textit{generalized} $U(1)_L$ charge assignment follows from the invariance of \equaref{model}, which should not be confused with the standard choice of the lepton numbers.

An electromagnetic anomaly to the $U(1)_L$ current necessarily arises regardless of the choices for $\cchi_{q}$, $\cchi_{1}$ and $\cchi_2$. The Lagrangian then gains the following term, allowing the decay of Majorons to photons:
\begin{equation}
    \mathcal{L} \supset  E\frac{J}{F_J}\frac{\alpha_{\mathrm{em}}}{4 \pi} F_{\mu \nu} \widetilde{F}^{\mu \nu},
    \label{jgammacoupling}
\end{equation}
where $E$ is the electromagnetic anomaly coefficient and $F_J$ is the Majoron decay constant. Given the dimension $d(C_i)$ of the color representation, the $U(1)_L$ charge $\mathcal{X}_i$, and the electromagnetic charge $Q_i$ of the fields, we can compute the anomaly coefficient of this model via the standard procedure,
\begin{equation}
    E = \sum_i d(C_i) \mathcal{X}_i Q_i^2,
\end{equation}
where all the fields are taken to be left handed so that $\mathcal{X}_{i, L} = - \mathcal{X}_{i, R}$. In particular, for $n_f$ generations of fermions, the model from \tabref{charges} yields
\begin{equation}
\begin{aligned}
    E &= n_f \left(\frac{5}{3} \mathcal{X}_q - \frac{4}{3}\mathcal{X}_u - \frac{1}{3}\mathcal{X}_d + \mathcal{X}_l - \mathcal{X}_e\right) \\
    &= n_f \left(\mathcal{X}_{2} - \mathcal{X}_{1} \right)~.
\end{aligned}
\label{EManomaly}
\end{equation}
Similarly, one can also show that this model does not have a QCD anomaly, regardless again of the choices for $\cchi_q, \cchi_1$ and $\cchi_2$. Here, we consider a general charge assignment for $\mathcal{X}_{2} - \mathcal{X}_{1} \equiv \cchi_{12}$. Cases of particular interest include $\cchi_{12} = 1~\mathrm{or}~2$, corresponding to the model in \refref{Liang:2024vnd} and the more general case $\cchi_{12}>2$. As we will see, we will generally be interested in larger electromagnetic anomalies in order to successfully explain direct collapse SMBH formation, and so we will focus on the case $\cchi_{12}>2$. In this latter case, the Higgs doublets are charged from a high dimensional operator of the form
\begin{equation}
    \frac{1}{\Lambda^{n-2}} H_2^{\dagger}H_1 \phi^n + \text { h.c. },
\end{equation}
where $\Lambda$ is an ultraviolet cutoff which is presumably at the string scale \cite{Arkani-Hamed:2006emk}. Following \equaref{EManomaly}, the electromagnetic anomaly and consequently, the Majoron decay rate, is therefore enhanced: 
\begin{equation}
    \Gamma_J = n_f^2 \cchi_{12}^2 \frac{\alpha_{\mathrm{em}}^2  m_J^3}{64 \pi^3 F_J^2}.
\end{equation}
As we will see, the formation of SMBHs by direct collapse, along with the potential detectability of Majorons in the infrared band, crucially depend on the enhancement of $\cchi_{12}$.

Without specifying $\cchi_{12}$, we can write down the generic $U(1)_L$ invariant scalar potential
\begin{equation}
\begin{aligned}
    V&\left(H_1, H_2, \phi\right) = \mu_1^2\left|H_1\right|^2+\mu_2^2\left|H_2\right|^2+\frac{\lambda_1}{2}\left|H_1\right|^4+\frac{\lambda_2}{2}\left|H_2\right|^4 \\
    &+\lambda_{12}\left|H_1\right|^2\left|H_2\right|^2 + \frac{\lambda_{\phi1}}{2}|\phi|^2\left|H_1\right|^2+\frac{\lambda_{\phi 2}}{2}|\phi|^2\left|H_2\right|^2\\
    &+\lambda_{\phi}\left(|\phi|^2-\frac{v_\phi^2}{2}\right)^2 + V_{\mathrm{mix}},
\end{aligned}
\label{scalarpotential}
\end{equation}
where $V_{\mathrm{mix}}$ is the additional mixing between $H_1, H_2$ and $\phi$ that depends on $\cchi_{12}$. In the `enhanced' case where $\cchi_{12} > 2$, this mixing term between $H_1$ and $H_2$ after the lepton number symmetry breaking becomes
\begin{equation}
    V_{\mathrm{mix}} \supset\frac{\lambda_{\phi 12} v_{\phi}^{n}}{2^{n/2} \Lambda^{n-2}} H_2^{\dagger}H_1 + \text{ h.c. }  \equiv \mu_{12}^2 H_2^{\dagger}H_1 + \text{ h.c. }.
\end{equation}
Notably, besides the high dimensional operator introduced by our charge assignment, the scalar potential in our model mimics the scalar potential of DFSZ axion models~\cite{Dine:1981rt, Zhitnitsky:1980tq, Bertolini:2014aia, Espriu:2015mfa, DiLuzio:2023ndz}, where the phenomenology of two Higgs doublet scenarios is already well-studied~\cite{Gunion:2002zf, Bernon:2015wef, DiLuzio:2023ndz}. 

In general, one of the CP-even scalar components of the Higgs (conventionally denoted as $h$ and $H$) needs to align with the direction of the electroweak vacuum in order to behave as the SM-like Higgs, avoiding constraints from the Large Hadron Collider (LHC). This alignment limit can be achieved in two ways. Firstly, we could have `alignment with decoupling,' where all the scalar degrees of freedom except $h$ becomes heavy and decouple from the low energy effective theory. Alternatively we could have `alignment without decoupling,' where a SM-like Higgs is obtained while allowing the rest of the scalar degrees of freedom to remain at low scales (including being even lighter than the electroweak scale) through proper choices of the parameters in \equaref{scalarpotential}. 

In our model, both alignment limits can be accomplished. Specifically, we rotate $(H_1, H_2)$ to the basis $(\Phi_1, \Phi_2)$ so that only $\Phi_1$ picks up the electroweak VEV. Then alignment without decoupling corresponds to the case where the coefficient of the term $\Phi_1^{\dagger}\Phi_1\Phi_1^{\dagger}\Phi_2$ approaches zero. The alignment with decoupling limit is more interesting, since it is closely connected to the mass mixing $\mu_{12}$, thus providing us a probe of the cutoff scale $\Lambda$. For instance, if we are interested in a $\cchi_{12} = 6$ model and if we have $v_\phi \simeq F_J \gtrsim 10^{11}$~GeV and $\Lambda \sim 10^{15}$~GeV, then $\mu_{12} \gtrsim \mathrm{TeV}$. This mixing scale also sets the scale of the non-SM-like scalars in the model to be within reach of the LHC and future colliders \cite{ATLAS:2024vxm, Beniwal:2022kyv, Arcadi:2022lpp, Dutta:2025nmy}. Due to the absence of FCNCs, our model is `flavor-safe' even when the non-SM-like scalars are at the TeV scale.

The additional couplings $|\phi|^2\left|H_1\right|^2$ and $|\phi|^2\left|H_2\right|^2$ in \equaref{scalarpotential} may spoil the hierarchy $v_{\phi} \gg v_{EW}$ if their coefficients are of $\mathcal{O}(1)$. However, it is argued in \refsref{Volkas:1988cm}{Foot:2013hna} that it is natural to set $\lambda_{\phi1}$ and $\lambda_{\phi2}$ to be small, since if they vanish, we obtain an enhanced Poincare symmetry for the model. Therefore, the Higgs sector remains radiatively stable even with the spontaneously broken $U(1)_L$ at a much higher scale\footnote{Without $V_{\mathrm{mix}}$, the scalar sector has a larger symmetry which corresponds to independent rotations of $H_1$, $H_2$ and $\phi$. If the theory is close to a conformal fixed point below the lepton number breaking scale, the two Higgs doublets are almost massless under this larger symmetry group. The hierarchy $v_{\phi} \gg v_{EW}$ can be equivalently maintained in this way, and the electroweak scale is set by $V_{\mathrm{mix}}$ where $\mu_{12} \sim \mathrm{TeV}$ can be achieved with suitable choices of $\Lambda$, $v_\phi$ and $n$.}.

\section{Majoron Dark Matter at eV Scales}
\label{Majorondm}

\noindent While the model we describe in \secref{Majoronmodel} does not contain a Majoron mass term, the global $U(1)_L$ symmetry is generally expected to be explicitly broken---for example, by quantum gravity effects \cite{Giddings:1987cg, Alonso:2017avz}. Hence, the Majoron gains a mass and becomes a pseudo Nambu-Goldstone boson. The possibility of the Majoron being the dominant component of dark matter is of particular interest not only because its relic density can be produced through either thermal freeze in/out or the misalignment mechanism in the early Universe, but also because the same model can generate the observed baryon asymmetry via leptogenesis \cite{Fukugita:1986hr, Buchmuller:2005eh}. 

The mass scale of the right handed neutrinos and thus the scale of leptogenesis is set by the parameter $F_J$. On the other hand, the masses of the active neutrinos is related to $F_J$ in the Type I seesaw mechanism by 
\begin{equation}
    m_{\nu} \sim Y_D \frac{v_{EW}^2 Y_N^{-1}}{F_J} Y_D^T.
\end{equation}
A well-motivated range of $F_J$ can be estimated by assuming that the Dirac mass of the heaviest neutrino is close to the mass of the $\tau$ lepton:
\begin{align}
    (Y_{D})_{33} v_{EW} \sim \mathrm{GeV}~.
\end{align}
Then, for $m_{\nu} \lesssim 0.01\ \mathrm{eV}$, we can restrict $F_J$ to,
\begin{align}
    F_J \gtrsim 10^{11} ~\mathrm{GeV}~.
\end{align}
Depending on the mass hierarchy of the Majorana masses, leptogenesis will happen at the scale around, or lower than, $10^{11} ~\mathrm{GeV}$, from the CP-violating out-of-equilibrium decay of the lightest right-handed neutrinos $N_1$. We also note that this choice of the parameter evades the lower bound on reheating temperature in leptogenesis \cite{Buchmuller:2004nz, Giudice:2003jh} and the Davidson-Ibarra bound on the lowest mass of $N_1$ \cite{Davidson:2002qv}.

Unlike previous works that focus on Majoron dark matter at keV or above \cite{Garcia-Cely:2017oco, Heeck:2017xbu, Bloch:2020uzh, deGiorgi:2023tvn, Akita:2023qiz}, we are interested in light Majorons with masses at the eV scale produced from the misalignment mechanism. In particular, we will assume a pre-inflationary scenario where the lepton number symmetry is broken sufficiently early before the end of inflation so that the initial misalignment angle $\theta_i$ is a free parameter. The Majoron abundance from the misalignment mechanism can be parametrized by \cite{OHare:2024nmr}
\begin{equation}
    \Omega_J h^2 \simeq 0.12\left(\frac{m_J}{25 \mathrm{eV}}\right)^{1 / 2}\left(\frac{F_J \theta_i}{2.7 \times 10^{11} \mathrm{GeV}}\right)^2,
\end{equation}
The Majoron dark matter works much the same as light axionic dark matter, forming a (non-thermal) condensate with the behavior of cold dark matter. We note that when Majoron comprise the dark matter, the previously-discussed natural choice of $F_J\gtrsim10^{11}$GeV corresponds nicely to Majorons in the eV mass range. Beyond the aforementioned theoretical motivations, the eV mass range is important also for practical reasons: 1) using existing telescope observations, we can probe a significant fraction of the parameter space for the Majoron-to-photon decay signals and constrain our Majoron models with $\cchi_{12} > 2$, and 2) for Majorons that can decay into Lyman-Werner photons, the supplied radiation background can be strong enough to trigger the formation of direct collapse black holes \cite{Lu:2024zwa}. Therefore, this eV mass window could connect the problem of early SMBH formation to the nature of neutrino mass and dark matter, while laying within reach of current instruments like the Hubble Space Telescope~\cite{Carenza:2023qxh}.

\begin{figure*}[ht!]
    \centering
    \includegraphics[width=.9\textwidth]{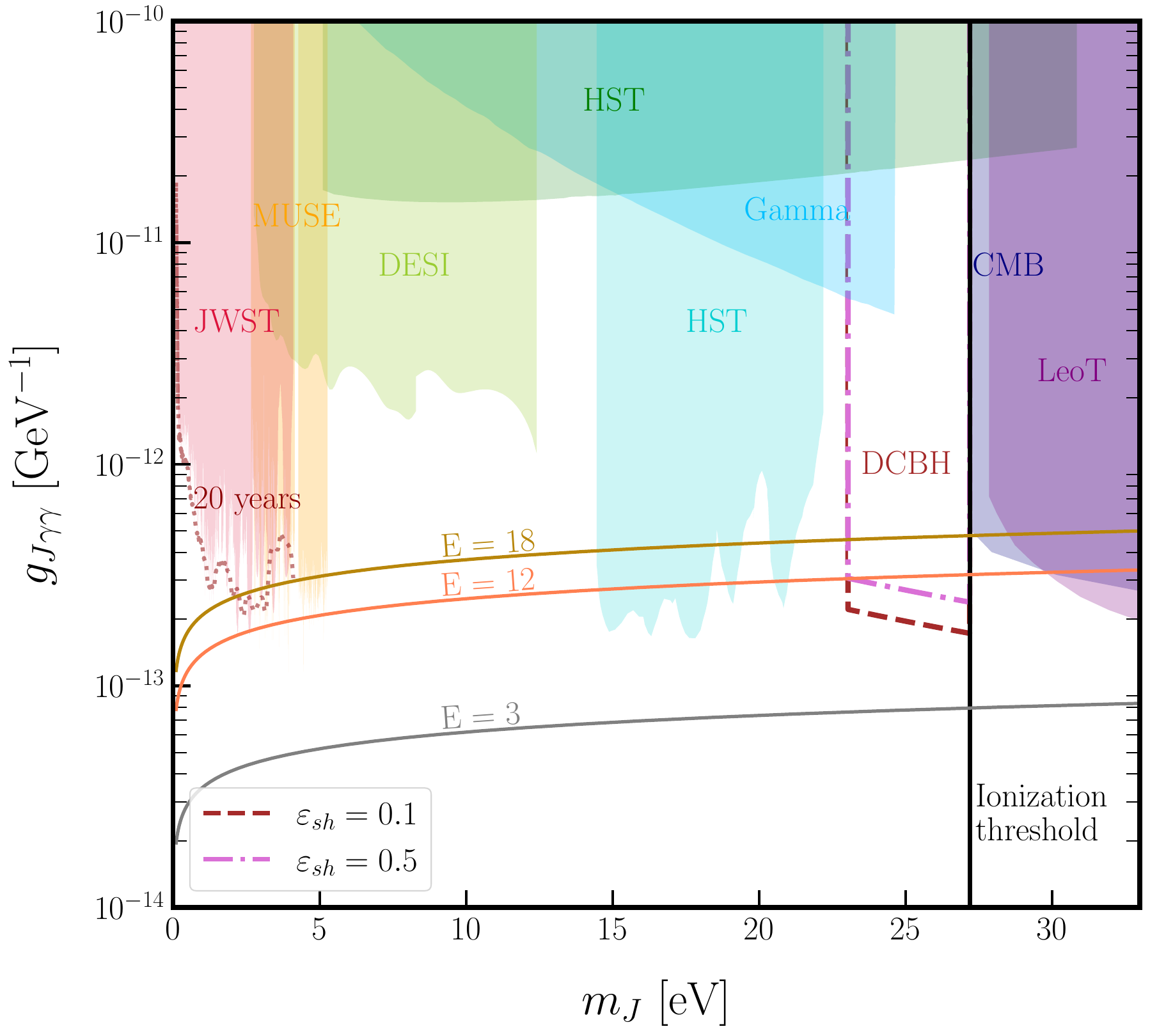}
    \caption{Parameter space and constraints for eV scale Majoron DM. Majoron models with different electromagnetic anomalies $E$ are plotted in solid lines, where we fix the Majoron abundance $\Omega_J h^2 = 0.12$ and choose $\theta_i = 3$. The dashed and dashdotted lines mark the region for direct collapse black holes formation. Current astrophysics constraints are shown in the color filled regions, while the projected end-of-mission reach of JWST is shown with the dotted line. The JWST projection is calculated using the smoothed version of the current constraint.}
    \label{fig:Majoron_para}
\end{figure*}

As an analogy to the axion-photon coupling, we can define the Majoron-photon coupling from \equaref{jgammacoupling} according to 
\begin{equation}
    g_{J\gamma\gamma} = \frac{\alpha_{\mathrm{em}}}{\pi} \frac{E}{F_J}.
\end{equation}
Notice that this Majoron-photon coupling is normalized in the same way as the standard axion-photon coupling. This allows us to directly compare $g_{J\gamma\gamma}$ with existing axion decay bounds.

In \figref{fig:Majoron_para}, we plot the relevant constraints on $g_{J\gamma\gamma}$ obtained from observations with JWST \cite{Pinetti:2025owq}, MUSE \cite{Todarello:2023hdk}, DESI \cite{Wang:2023imi}, and HST \cite{Todarello:2024qci, Carenza:2023qxh}, together with constraints from gamma-ray attenuation (Gamma) \cite{Bernal:2022xyi}, CMB anisotropy (CMB) \cite{Bolliet:2020ofj}, and the heating of dwarf galaxies (LeoT) \cite{Wadekar:2021qae}. For reference, three models with different anomaly coefficients are also plotted where $E = 3$ is the model in \refref{Liang:2024vnd}, $E = 12$ corresponds to a model with a dimension six operator ($n = 4$), and $E = 18$ (or $n = 6$) is the model discussed in \secref{Majoronmodel}. 

It is worth noting that the JWST blank-sky measurements contribute to a great advancement in constraining dark matter decay signals in the infrared range, corresponding to Majoron mass within $0.5-4~\mathrm{eV}$. Furthermore, the constraints obtained from \refref{Pinetti:2025owq} are stronger than those in the previous analysis \cite{Janish:2023kvi, Roy:2023omw, Saha:2025any} by around two orders of magnitude due to the selection of targets with a higher decay signal and longer than expected integration times. Consequently, our $E = 18$ model is reached even with the current 2.5 years of data. In \refref{Pinetti:2025owq}, the constraints are obtained using the $\chi^2$ statistics constructed from the blank-sky data. Therefore, the statistics scales linearly with the total integration time and is proportional to $g_{J\gamma\gamma}^4$. In Ref.~\cite{Janish:2023kvi}, estimates of the future reach of JWST were obtained by scaling the current constraints with a factor $(t_{\text{tot}}/\mathrm{2.5~yr})^{-1/4}$, assuming a constant fraction of integration time throughout the mission. Taking $t_{\text{tot}} = \mathrm{20~yr}$ (the expected lifespan of JWST), the reach of JWST at the end of the mission can then be estimated, with the projected constraints plotted in \figref{fig:Majoron_para}. Furthermore, a dedicated observation of high dark matter density regions, such as the Galactic Center, can significantly strengthen the projected constraints, bringing even the $E = 12$ model within reach.

\section{SMBH from Majoron Decay}
\label{SMBH}

\noindent One of the most striking discovery from the JWST is the surprising abundance of active galactic nuclei (AGNs) powered by SMBHs at high redshifts \cite{Bogdan:2023ilu, ubler2023ga, larson2023ceers, harikane2023jwst, carnall2023massive, onoue2023candidate, kocevski2023hidden, Maiolino:2023zdu, fan2023quasars}. The existence of this population of seemingly overmassive black holes challenges our understanding of their origin and growth history. Producing such early SMBHs from `light seeds' (e.g. the remnants of Pop III stars) requires extended periods of super-Eddington accretion \cite{ohkubo2009evolution, Banik:2016qww, kroupa2020very}, but even assuming this is possible, feedback events could stall subsequent accretion \cite{huvsko2025effects, Massonneau:2022sye} and it remains challenging to rapidly grow SMBHs from this route. On the other hand, direct collapse black holes \cite{Omukai:2000ic, Oh:2001ex, Begelman:2006db, Inayoshi:2019fun, Bromm:2002hb, Choi:2013kia, shang2010supermassive, Dijkstra:2008jk, becerra2018assembly} can provide heavy seeds for SMBH growth, being born as black holes at masses closer to $10^5 M_{\odot}$. Furthermore, as estimated in \refref{Friedlander:2022ovf}, the comoving number density of the eligible halos for direct collapse has the potential to explain the observed high redshift quasar densities. Although either seed scenarios cannot be excluded by the current JWST data \cite{Volonteri:2022yhe, Schneider:2023xxr}, SMBHs discovered in some extreme environments \cite{ono2022morphologies, Natarajan:2023rxq, Kovacs:2024zfh} may favor the direct collapse scenario.

The key for the successful formation of direct collapse black holes is the suppression of $\htwo$ formation in  pristine, metal-free hydrogen clouds. Sufficient suppression of molecular Hydrogen prevents catastrophic cooling and the fragmentation of the cloud. Consequently, Pop III star formation is suppressed and the cloud can collapse monolithically into an unstable supermassive star and ultimately collapsing into an intermediate mass black hole. 

In general, there are two approaches to suppressing the formation of molecular Hydrogen in the clouds: 1) providing a Lyman-Werner radiation background ($11.2 - 13.6~\mathrm{eV}$) can effectively dissociate $\htwo$ and detach electrons from $\hminus$, which disrupts the major channel of $\htwo$ formation~\cite{Bromm:2002hb, shang2010supermassive, Choi:2013kia, Cyr:2022urs, Friedlander:2022ovf, Lu:2024zwa}, or 2) introducing an additional heating source that raises the cloud temperature up to $10^4$ K before the period of collapse, leading to the destruction of $\htwo$ through collisional dissociation \cite{sethi2010supermassive, Lu:2023xoi}. We will focus on the first approach in this section as the decay of Majorons provides a natural radiation source in the Lyman-Werner band.

In the literature, a common source of Lyman-Werner radiation is nearby star forming galaxies \cite{Omukai:2000ic,Oh:2001ex,Dijkstra:2008jk,shang2010supermassive}. The problem of determining the critical radiation intensity for this constant, external background has been studied using one-zone models and hydrodynamical simulations \cite{agarwal2016new, Wolcott-Green:2016grm, luo2020direct}. However, it is unclear if the abundance of the SMBHs produced in this way would be high enough to match the observed AGN densities, since the radiation only comes from the tails of the Pop III (or Pop II) star spectra. Additionally, when the column density of $\htwo$ becomes high enough so that the cloud is self-shielding, it remains challenging to reliably estimate the effective photodissociation rate $k_{\mathrm{H}_2}$. A common procedure is to parametrize the effective rate by an additional factor $f_{\text {shield }}$ \cite{Draine:1996hna}, so that the full dissociation rate becomes
\begin{equation}
    k_{\mathrm{H}_2}\left(N_{\mathrm{H}_2}, T\right)=k_{\mathrm{H}_2} f_{\text {shield }}\left(N_{\mathrm{H}_2}, T\right),
\end{equation}
where $f_{\text {shield }}$ depends both on the $\htwo$ column density $N_{\mathrm{H}_2}$, and the gas temperature $T$.

On the other hand, the decay of Majorons generates an \textit{in-situ} radiation source that not only reduces the self-shielding, but also becomes ubiquitous to all of the atomic cooling halos that are able to directly collapse. As the radiation intensity depends on the dark matter density, which evolves dynamically during the collapse, we use a one-zone framework developed in \refsref{Lu:2023xoi}{Lu:2024zwa} to track the co-evolution of the baryon cloud and the adiabatic contraction of the dark matter halo. In this one-zone model, the redshift-dependent specific intensity (in units of $\mathrm{J/cm^2/s/Hz/sr}$) can be calculated as
\begin{equation}
    J(E, z)= \Gamma_J \frac{f_J(z) \bar{\rho}_{D M}(z) R_h}{m_J/2}\delta\left(1-\frac{2 E}{m_J}\right),
\end{equation}
where we keep track of the averaged dark matter density $\bar{\rho}_{D M}(z)$, the halo radius $R_h$, and the remaining dark matter fraction $f_J(z)$ due to the decay. We also note that this monochromatic spectrum resulting from the two body decay has a significantly different spectral energy distribution from the typical Pop III/Pop II type spectra used in direct collapse simulations. The photondissociation and photodetachment rates are related to the specific intensity by \cite{Wolcott-Green:2016grm}
\begin{equation}
\begin{aligned}
& k_{\mathrm{H}^{-}}(z)=\int_{0.76 \mathrm{eV}}^{13.6 \mathrm{eV}} 4 \pi \sigma_{\mathrm{H}^{-}}(E) \frac{J(E, z)}{E} \frac{d E}{h}, \\
& k_{\mathrm{H}_2}(z) \approx 1.39 \times 10^{-12} \mathrm{~s}^{-1} \frac{J_{L W}(z)}{J_{21}}.
\end{aligned}
\end{equation}
In the above reaction rates, we use the cross section $\sigma_{\mathrm{H}^{-}}$ in \refref{shapiro1987hydrogen} and $J_{L W}$ is the averaged intensity in the Lyman-Werner band. As the reduction in the cloud self-shielding cannot be straightforwardly modeled in our one-zone approach, we introduce the parameter $\varepsilon_{sh}$ to account for the reduction
\begin{equation}
    f_{\text {in-situ }}=1-\varepsilon_{s h}\left(1-f_{\text {shield }}\right).
\end{equation}
Due to the in-situ production of radiation, the self-shielding effect should be greatly suppressed and we expect $\varepsilon_{s h} \ll 1$. For a more generic decaying dark matter model, the effect of varying $\varepsilon_{s h}$ was explicitly computed in Ref.~\cite{Lu:2024zwa}. For our more fully realized model here, we use the same chemical network as in \refref{Lu:2024zwa} to track the change in temperature and chemical components during the collapse. A successful direct collapse process is identified by a final temperature around $10^4~\text K$ and an $\htwo$ fraction $\ll 10^{-5}$.

In \figref{fig:Majoron_para}, we plot the region where the Majoron decay can produce enough radiation to trigger the direct collapse process (labeled `DCBH') with different values of $\varepsilon_{sh}$. Notably, with a reasonable choice of $\theta_i$, both of our $E = 18$ and $E = 12$ models fall inside the relevant region for direct collapse, allowing us to simultaneously explain the mystery of dark matter, neutrino masses, the baryon asymmetry of the Universe, and high redshift SMBHs.

As a final note, our one-zone approach can indeed capture many crucial features in the direct collapse mechanism, such as the suppression of $\htwo$ formation and the high cloud temperature under the atomic cooling phase. However, the need for a full hydrodynamical simulation cannot be over-emphasized: it provides a complete perspective in understanding the complex history of SMBH growth and its interaction with the gas environments \cite{Jeon:2024iml}, and motivates diagnostics that distinguishes different seed scenarios \cite{nakajima2022diagnostics}. More importantly, our direct collapse model is devoid of the commonly used Lyman-Werner background, potentially leading to different metal enrichment signatures that could be tested with simulations. There are currently a wide range of high redshift AGNs that lack a comprehensive explanation, such as the one discussed in \refref{Maiolino:2025tih}, and our model could be applicable to a broad class of early SMBHs in rare environments.

\section{Conclusion}
\label{conclusion}

\noindent The light Majoron as a dark matter candidate is a topic that remains relatively unexplored. In this work, we studied two Higgs doublet Majoron models with an enhanced electromagnetic anomaly, where a dark matter relic abundance is generated through the misalignment mechanism. The spontaneous breaking of the lepton number symmetry gives a large Majorana mass to the right-handed neutrinos, which makes Standard Model neutrinos naturally light, while the decay of the heavy neutrinos produces a baryon asymmetry via leptogenesis. From considerations regarding the neutrino masses, the misalignment abundance, and astrophysical detectability, we demonstrated that perhaps the best motivated parameter range for the Majoron lies in the eV window. Interestingly, this mass window is also intimately connected to the formation of high redshift SMBHs in direct collapse scenarios. With an appropriately enhanced anomaly, the Majoron decay signals are within reach of our current generation of instruments. In the Higgs sector, the high dimensional operator responsible for the anomaly enhancement also allows for additional Higgs bosons near TeV that can be detected in future high energy experiments.

Our model can be generalized in many interesting directions. One possibility is to include a more complex dark sector with fermions carrying the lepton number. Anomaly enhancement could then be achieved with a proper lepton charge assignment without the need for high dimensional operators. We leave the exploration of this direction for future work.

JWST observations have so far greatly advanced our understanding of the conditions in the early Universe, providing state-of-the-art constraints on Majoron decay in the infrared mass range. The current constraints could be significantly strengthened by devoting more observation time to the Galactic Center, as the D-factor is improved by more than one order of magnitude compared to other targets used in the setting of these constraints \cite{Pinetti:2025owq}.

Finally, we demonstrated the intriguing relation between the `ultraviolet' Majoron dark matter and the possibility of direct collapse SMBH formation. Our one-zone model motivates a full-scale simulation capable of uncovering essential properties such as morphology, metal enrichment, and the black hole–star mass relation. We demonstrate here again that unraveling the origin of high-redshift SMBHs may eventually elucidate the nature of dark matter as well.

\section*{Acknowledgements}
\noindent We thank Elena Pinetti and Qiuyue Liang for helpful discussions.
The work of Y.L., Z.P. and A.K. was supported by the U.S. Department of Energy (DOE) Grant No. DE-SC0009937. A.K. was also supported by World Premier International Research Center Initiative (WPI), MEXT, Japan; and by the Japan Society for the Promotion of Science (JSPS) KAKENHI Grant No. JP20H05853. The work of T.T.Y. was supported by the
Japan Society for the Promotion of Science (JSPS) KAKENHI Grant No. JP24H02244. This work made use of N\textsc{um}P\textsc{y}~\cite{numpy2020Natur.585..357H}, S\textsc{ci}P\textsc{y}~\cite{scipy2020NatMe..17..261V}, and M\textsc{atplotlib}~\cite{mpl4160265}.

\bibliography{bib.bib}
\bibliographystyle{bibi}

\end{document}